\documentstyle[aps,pre]{revtex}
\catcode`\@=11

\def\maketitle2{\par 
\begingroup
\let\cite\@bylinecite
\def\thefootnote{\fnsymbol{footnote}}
\twocolumn[\@maketitle2\vskip2pc] \thispagestyle{plain}\@thanks
\endgroup

\def\thefootnote{\arabic{footnote}}
\setcounter{footnote}{0}
\let\maketitle2\relax \let\@maketitle2\relax
\let\@thanks\relax \let\@authoraddress\relax \let\@title\relax
\let\@date\relax \let\thanks\relax \let\@abstract\relax
\let\@pacs\relax}

\def\abstract#1{\gdef\@abstract{{\par 
\bgroup \ifdim\prevdepth=-1000pt \prevdepth0pt\fi
\hsize\columnwidth \dimen0=-\prevdepth \advance\dimen0 by17.5pt
\nointerlineskip \small\vrule width 0pt height\dimen0
\relax}{~~}#1\egroup}}

\def\pacs#1{\gdef\@pacs{{\par 
\bgroup \hsize\columnwidth \parindent0pt \ifdim\prevdepth=-1000pt
\prevdepth0pt\fi \dimen0=-\prevdepth \advance\dimen0
by20pt\nointerlineskip \egroup} PACS numbers:~#1}}

\def\@maketitle2{
\@preprint \@title \ifdim\prevdepth=-1000pt \prevdepth0pt\fi
\@authoraddress \@date
\begin{list}{}{\leftmargin=0.10753\textwidth \rightmargin=\leftmargin
\itemsep=1pc\partopsep=-1pc}
\item\@abstract
\item\@pacs
\end{list}}

\catcode`\@=12

\begin{document}
\title{Magnetization in quasiperiodic magnetic multilayers with biquadratic
 exchange coupling}

\author{C.G. Bezerra}\address{Departamento de F\'\i sica,
Universidade Federal do Rio Grande do Norte,\\ 59072-970,
Natal-RN, Brazil}
\author {J. M. de Ara\'ujo}\address{Departamento de Ci\^encias Naturais, Universidade
Estadual do Rio Grande do Norte,\\ 59610-210, Mossor\'o-RN,
Brazil}
\author{C. Chesman and E. L. Albuquerque\thanks {Corresponding author,
e-mail: ela@dfte.ufrn.br}}\address{Departamento de F\'\i sica,
Universidade Federal do Rio Grande do Norte, \\ 59072-970,
Natal-RN, Brazil}

\date{\today}

\abstract {\small{ A theoretical study of the magnetization curves
of quasiperiodic magnetic multilayers is presented. We consider
structures composed by ferromagnetic films (Fe) with interfilm
exchange coupling provided by intervening nonferromagnetic layers
(Cr). The theory is based on a realistic phenomenological model,
which includes the following contributions to the free magnetic
energy: Zeeman, cubic anisotropy, bilinear and biquadratic
exchange energies. The experimental parameters used here are based
on experimental data recently reported, which contain sufficiently
strong biquadratic exchange coupling.}}


\maketitle2 \narrowtext

\section{Introduction}
The study of the properties of magnetic multilayers has been one
of the most investigated fields in the last decade. The
understanding of a number of new and intriguing results became an
exciting challenge from both theoretical and experimental point of
view. In their pioneer work, Gr\"unberg and colaborators\cite{1}
reported evidences of an antiferromagnetic bilinear exchange
coupling in Fe/Cr/Fe structures. After that, Baibich {\it et
al}\cite{2} noticed a sudden fall in the electrical resistance of
Fe/Cr magnetic multilayers when an external magnetic field was
applied. The effect was so striking that was called giant
magnetoresistance, and recently it has been widely considered for
applications in information storage technology\cite{3}. Through
magnetoresistance measurements, Parkin, More and Roche\cite{4}
observed an oscillatory behavior of the exchange coupling in
magnetic metallic multilayers as a function of the nonmagnetic
spacer thickness. This work was seminal to a number of
experimental studies on Fe/Cr/Fe structures with different
nonmagnetic spacer thickness. Later on, in 1991, R\"uhrig {\it et
al}\cite{5} showed evidences of a non-colinear alignment
$(90^{\circ})$ between ferromagnetic layers in Fe/Cr magnetic
multilayers, for nonmagnetic spacer thickness, where the bilinear
exchange coupling was small. This behavior could not be explained
considering only the usual bilinear exchange coupling in the free
magnetic energy. In fact, the inclusion of a biquadratic exchange
term in the free magnetic energy of the system allows the
stabilization of non-colinear alignments. Until recently, it was
found that the biquadratic exchange coupling was too small when
compared to the bilinear exchange coupling. However, Azevedo {\it
et al}\cite{6,7,8} presented a number of experimental results in
Fe/Cr/Fe samples which show the biquadratic exchange coupling
comparable to the bilinear exchange coupling. Therefore, the
biquadratic coupling can play a remarkable role in the properties
of magnetic multilayers.

On the other hand, from an experimental point of view, due to the
rapid development of the crystal growth techniques, is now
possible to tailor a wide class of magnetic multilayers, whose
film thickness is extremely well controlled. As a consequence,
there are new magnetic phases and properties which are not shared
by the constituent materials.

It is known that the magnetic properties can depend strongly on
the stacking pattern of the layers. Under this aspect, the
physical properties of a new class of artificial material, the
so-called quasiperiodic structures, became recently an attractive
field of research. Quasiperiodic structures, which can be
idealized as the experimental realization of a one-dimensional
quasicrystal, are composed by the superposition of two (or more)
building blocks that are arranged in a desired manner. They can be
defined as an intermediate state between an ordered system (a
periodic crystal) and a disordered one (an amorphous
solid)\cite{9,10}. One of the most interesting features of these
systems is that the long range correlations, induced by the
construction of the systems, are reflected in their various
spectra. In fact, many physical properties of quasiperiodic
systems have been studied such as light propagation\cite{11},
phonons\cite{12}, electronic transmission\cite{13},
polaritons\cite{14}, and magnons\cite{15} . In all of these
situations, despite the diversity of the systems, a common feature
is present, namely, a fractal spectra of energy, which can be
considered as their basic signature\cite{16,17}. However only very
recently some efforts were done towards the understanding of the
properties of quasiperiodic magnetic multilayers\cite{15,18}.

The main aim of this paper is a contribution to the understanding
of the effects of the quasiperiodic arrangement on the
magnetization curves in magnetic multilayers. We are interested in
new magnetic phases and alignments that are only due to the
quasiperiodicity of the system. We have studied Fe/Cr$(100)$
structures which follow a Fibonacci and a double period or {\it
 generalized Fibonacci} quasiperiodic sequences.

The layout of the paper is as follow: In section II we discuss the
physical model used here, with emphasis in the description of the
quasiperiodic sequences. In section III we define the
contributions to the magnetic energy. The numerical methods, used
to obtain the equilibrium configuration, are described in section
IV. In section V, the results are presented and discussed.
Finally, we draw the conclusions in section VI.

\section{Physical model}
A quasiperiodic structure can be experimentally constructed
juxtaposing two building blocks  (or, as considered here, building
layers) following a given quasiperiodic sequence. We choose Fe as
the building layer associated with the letter $A$, and Cr as the
building layer associated with the letter $B$ (see Fig.\ 1).
Therefore, we only take into account generation of sequences that
start and finish with an Fe building layer, which means an even
number of Fe layers, to guarantee a real magnetic counterpart. In
this way we avoid also the intriguing behavior found when even and
odd numbers of Fe layers are considered, as discussed in
\cite{19}. In this paper we have considered two quasiperiodic
sequences, namely, the Fibonacci and the double period sequences.

\subsection{The Fibonacci magnetic multilayers}
The $N$th generation of the Fibonacci sequence can be determined
appending the $N-2$ generation to the $N-1$ one, i.e.,
$S_{N}=S_{N-1}S_{N-2}$ ($N \geq 2$). This algorithm construction
requires initial conditions which are chosen to be $S_{0}=B$ and
$S_{1}=A$. The Fibonacci generations can also be alternatively
obtained by an iterative process from the substitution rules (or
inflation rules), $A \rightarrow AB$, $B \rightarrow A$. The
Fibonacci generations are:

\begin{center}
$S_0=\left[ B \right]$  , $S_1=\left[ A \right]$  ,
$S_2=\left[A  B \right]$  , $S_3=\left[ AB  A \right]$ \ \ etc.
\end{center}

\noindent In a given generation $S_{N}$, the total number of
letters is given by the Fibonacci number $F_{N}$, which is
obtained by the relation $F_{N}=F_{N-1}+F_{N-2}$, with
$F_{0}=F_{1}=1$. Also, $F_{N-1}$ and $F_{N-2}$ are the number of
letters $A$ and $B$, respectively. As the generation order
increases ($N
>> 1$), the ratio $F_{N}/F_{N-1}$ approach to $\tau = (1+
\sqrt5)/2$, an irrational number which is known as the golden
mean. It is also possible to obtain the number of letters $A$ and
$B$ for a given generation by the substitution matrix of the
Fibonacci sequence ${\mathbf M_{F}}$ from\cite{20},

\begin{equation}
\left[  \begin{array}{c}  n_{A}^{N+1} \\ \\ n_{B}^{N+1} \\ \end{array}
 \right]
={\mathbf M_{F}}\left[  \begin{array}{c}  n_{A}^{N} \\ \\ n_{B}^{N} \\
\end{array}\right].
\end{equation}
Here $(n_{A}^{N+1},n_{B}^{N+1})$ are the number of letters $A$ and
$B$ in the $(N+1)th$ generation, and $(n_{A}^{N},n_{B}^{N})$ are
the number of letters $A$ and $B$ in the $Nth$ generation. The
explicit form of the substitution matrix for the Fibonacci
sequence is,

\begin{equation}
{\mathbf M_{F}}=\left[  \begin{array}{cc} 1 & 1\cr
1 & 0\end{array} \right],
\end{equation}
\noindent whose first eigenvalue, $\lambda$, is the golden mean
$\tau$.

In  Fig.\ 1, we show the third and fifth Fibonacci generations and
their magnetic counterparts. Note that the third Fibonacci
generation corresponds to a trilayer Fe/Cr/Fe, and in the fifth
Fibonacci generation there is a double Fe layer. It is easy to
show that the Fibonacci magnetic multilayers, for any generation,
are composed by single Cr layers, single Fe layers and double Fe
layers. The number of Fe single layers is $1+F_{N-2}$, the number
of Fe double layers is $-1+ F_{N-1} - F_{N-2}$ and the number of
Cr layers is $F_{N-2}$. It should be observed that only odd
Fibonacci generations have a magnetic counterpart (they start and
finish with an Fe building layer).

\subsection{The double period magnetic multilayers}
The N$th$ generation of the double period sequence can be obtained
from the relations,
\begin{equation}
S_{N}=S_{N-1}S^{\dag}_{N-1},
\end{equation}
with
\begin{equation}
S^{\dag}_{N}=S_{N-1}S_{N-1} \ \ (N\geq2),
\end{equation}
\noindent The initial conditions are $S_{0}=A$ e $S_{1}=AB$. We
can alternatively use the substitution rules $A \rightarrow AB$,
$B \rightarrow AA$. The  double period generations are:

\begin{center}
$S_0=\left[ A \right]$  , $S_1=\left[ AB \right]$  ,
$S_2=\left[A  BAA \right]$  , \ etc.
\end{center}

\noindent In a given generation $S_{N}$, the total number of
letters is $2^{N}$, and the number of letters $A$ and $B$ for
consecutive generations can be related by the substitution matrix
of the double period sequence ${\mathbf M_{dp}}$, i.e.:\cite{20}

\begin{equation}
\left[  \begin{array}{c}  n_{A}^{N+1} \\ \\ n_{B}^{N+1} \\ \end{array}
 \right]
={\mathbf M_{dp}}\left[  \begin{array}{c}  n_{A}^{N} \\ \\ n_{B}^{N} \\
\end{array}
\right].
\end{equation}
Here $(n_{A}^{N+1},n_{B}^{N+1})$ are the number of letters $A$ and
$B$ in the $(N+1)th$ generation, and $(n_{A}^{N},n_{B}^{N})$ are
the number of letters $A$ and $B$ in the $Nth$ generation. As the
generation number increases ($N>>1$), the ratio between the number
of letters $A$ and $B$ tends to $2$. The explicit form of the
substitution matrix for the double period sequence is,
\begin{equation}
{\mathbf M_{F}}=\left[  \begin{array}{cc} 1 & 1\cr
2 & 1\end{array} \right].
\end{equation}

In Fig.\ 2 we show the second and fourth double period generations
and their magnetic counterparts. The double period magnetic
multilayers are composed by single Fe layers, double Fe layers,
triple Fe layers and single Cr layers. It should be observed that,
contrary to the Fibonacci case, only even double period
generations have a magnetic counterpart.

\section{Magnetic energy}
We consider magnetic multilayers whose constituents are Fe
ferromagnetic films, separated by Cr non-magnetic films. We take
the $xy$-plane as the film plane and the $z$-axis as the growth
direction. We consider that the magnetic films are uniformly
magnetized and that they behave as monodomains. We also consider
that they do not present dynamical excitations and that the very
strong demagnetization field, generates by tipping the
magnetization out of the plane, will suppress any tendency for the
magnetization to tilt out of plane. Therefore, the degrees of
freedom of the magnetizations are restricted to the $xy$-plane.
The interfilm exchange couplings between the ferromagnetic films
are weak when compared to the strong exchange couplings between
spins within a given ferromagnetic film. Therefore, we can
represent the ferromagnetic films as classical magnetizations
$\vec M$, composed by the real spins within the films, which are
strongly coupled by the intrafilm exchange coupling. These
classical magnetizations interact through the interfilm exchange
coupling and they can present some anisotropy depending on the
structure studied. It should be noted that this system is
isomorphous to a one-dimensional chain of classical spins.

The global behavior of this system is well described by a
realistic phenomenological theory in terms of the free magnetic
energy\cite{7}, i.e.,
\begin{equation}
E_{T}=E_{z}+E_{ca}+E_{bl}+E_{bq}.
\end{equation}
Here $E_{z}$ is the Zeeman energy (between the ferromagnetic films
and the external applied magnetic field), $E_{ca}$ is the cubic
crystalline anisotropy energy (which we consider present in the
ferromagnetic films) and $E_{bl}$ and $E_{bq}$ are the bilinear
and the biquadratic exchange coupling energies (between the
ferromagnetic films), respectively.

The explicit form of the free magnetic energy can be written as,
\begin{eqnarray}
\lefteqn{ E_T=-\sum_{i=1}^n t_i \vec{M_i} \cdot \vec{H}} \nonumber
\\ &&+ \sum_{i=1}^n {\frac {t_iK_{ca}} {{|M_i |}^4} }(M_{ix}^2
M_{iy}^2 + M_{ix}^2 M_{iz}^2 + M_{iy}^2 M_{iz}^2) \nonumber \\
&&-\sum_{i=1}^{n-1}J_{bl}{\frac {\vec{M_i}\cdot\vec{M}_{i+1}}
{|\vec{M}_i||\vec{M}_{i+1}|} } +\sum_{i=1}^{n-1}J_{bq} {\frac
{{(\vec{M}_i\cdot\vec{M}_{i+1})}^2}
{|\vec{M}_i|^2|\vec{M}_{i+1}|^2}}.
\end{eqnarray}
Here, $\vec{H}$ is the external magnetic field which is applied in
the film plane, $t_i$ is the thickness of the $ith$ Fe layer,
$\vec{M_i}$ is the classical magnetization of the $ith$ Fe layer,
and $K_{ca}$ is the cubic anisotropy constant. Also, $J_{bl}$ and
$J_{bq}$ are the bilinear and the biquadratic exchange couplings,
respectively. This expression, after a tedious but straight
calculation, takes the form,
\begin{eqnarray}
\lefteqn
{E_{T}=\sum_{i=1}^n{\{-t_{i}M_{i}H\cos(\theta_{i}-\theta_H) + {1
\over 4}t_{i}K_{ca}{{\sin}^2{(2\theta_{i})}}\}}} \nonumber \\ &&
+\sum_{i=1}^{n-1}{\{-J_{bl}\cos(\theta_{i}-\theta_{i+1})
+J_{bq}{{\cos}^2{ (\theta_{i}-\theta_{i+1})}}\}}.
\end{eqnarray}
Here $\theta_{i}$ is the angular orientation of the magnetization
of the $ith$ Fe layer and $\theta_{H}$ is the angular orientation
of the magnetic field. From this point we consider $\theta_{H}=0$,
which means that the magnetic field is applied along the easy
axis. It is usual to write the total free magnetic energy in terms
of experimental parameters, like
\begin{eqnarray}
H_{ca}={\frac {2K_{ca}} {M_{S}}}, \\ H_{bl}={\frac {J_{bl}}
{tM_{S}}},  \\ H_{bq}={\frac {J_{bq}} {tM_{S}}}.
\end{eqnarray}
In this way we obtain a final expression for the free magnetic energy
per unit area,
\begin{eqnarray}
\lefteqn {{{E_{T}} \over {tM_{S}}}=\sum_{i=1}^n(t_{i}/t){\{-H_{0}\
cos(\theta_{i})+ {1 \over 8}H_{ca}{{\sin}^2{(2 \theta_{i})}}\}}}
\nonumber
\\&& +\sum_{i=1}^{n-1}{\{-H_{bl}\cos(\theta_{i}-\theta_{i+1})
+H_{bq}{{\cos}^2{ (\theta_{i}-\theta_{i+1})}}\}}.
\end{eqnarray}
Here $t$ is the thickness of a single Fe layer which is considered
to be the basic tile, $M_{i}$ is assumed to be equal to $M_{S}$
(the saturation magnetization), and $H_{ca}$ is the cubic
anisotropy field which turns the $(100)$ direction an easy
direction. $H_{bl}$ is the bilinear exchange coupling field which
favors antiferromagnetic alignment when negative, and
ferromagnetic alignment when positive. $H_{bq}$ is the biquadratic
exchange coupling field which is experimentally found to be
positive and favors a non-colinear alignment ($90^{\circ}$)
between two adjacent magnetizations.

Once the free magnetic energy is determined, we can calculate the
equilibrium configuration for specific values of the experimental
parameters as a function of the external applied field. In simple
situations, the equilibrium configuration can be analytically
obtained by equating to zero the derivatives of the magnetic
energy with respect to the angle $\theta$. However, in most cases
this leads to transcendental equations which can not be
analytically solved. From a numerical point of view, many methods
have been proposed to calculate the equilibrium positions of the
magnetizations. In the next section we describe the methods used
in this paper.

\section{Numerical methods}
In this section, we want to find the global minimum of the cost
function,
\begin{equation}
E_T=E_T(\theta_1,\theta_2,\ldots,\theta_n).
\end{equation}
where $\theta_n$ can assume values in the range $[0,2\pi]$ and it
defines a n-dimensional space. When the dimension of this space is
high, the cost function has a rough surface, i.e., there are many
local minima which make difficult to find the global minimum.
There are many numerical methods to solve this problem\cite{21}.
In our specific case two methods were successfully used, namely,
simulated annealing and the so-called gradient method.

\subsection{Simulated annealing method}
Introduced by S. Kirkpatrick\cite{22}, simulated annealing (SA)
comes from the fact that the heating (annealing) and slowly
cooling a metal, brings it into a more uniforme crystalline state,
which is believed to be the state where the free energy of bulk
matter takes its global minimum. The role played by the
temperature is to allow the configurations to reach higher energy
states with probability given by Boltzmann's exponential law. Then
they can overcome energy barriers that would otherwise force them
into local minima. In general, a simulated annealing technique can
be written as follows:

\begin{enumerate}

\item[(i)] Choose an initial point in parameter space, corresponding
to an initial configuration $\{\theta\}_j$, and calculate the
associated energy $E_j$.

\item[(ii)] Choose a second point in parameter space, corresponding to
a second configuration $\{\theta\}_{j+1}$, and calculate the
associated energy $E_{j+1}$.

\item[(iii)] If $\Delta E= E_{j+1} -E_j <0$, $\{\theta\}_{j+1}$ is
the new configuration of the system.

\item[(iv)] If $\Delta E\ge 0$, we define the probability
$p=\exp (-\Delta E/k_BT)$ and choose a random number $0\le x\le
1$. If $x\ge p$, $\{\theta\}_{j+1}$ is the new configuration of
the system. Otherwise, $\{\theta\}_j$ is maintained as the
configuration of the system.

\item[(v)] This procedure is executed again and again until the
equilibrium is reached.

\end{enumerate}

\subsection{The gradient method}
The second method that we have used was the so-called gradient
method\cite{23}. This method is based on the directional
derivative of the cost function (the magnetic energy) in the
search of its global minimum. In this way, we need to calculate
the gradient of $E_T$ with relation to the set $\{\theta\}$,
\begin{equation}
\vec{\nabla}E_T = \sum_{i=1}^n {\partial E_T\over
\partial\theta_i} \hat{\theta}_i.
\end{equation}

\noindent From this relation we execute the following algorithm to
find the equilibrium configuration,
\begin{enumerate}

\item[(i)] We generate a configuration in the parameter space $\{\theta\}_j$
from which we calculate the associated energy $E_j$ and the
gradient of the cost function.

\item[(ii)] A second point in the parameter space is generated by
$\{\theta\}_{j+1} =\{\theta\}_j - \alpha\vec{\nabla}E_T$. Here
$\alpha$ controls the size of the displacement in the direction
$-\vec{\nabla}E_T$.

\item[(iii)] The energy of the second point is calculated and if
$E_{j+1}>E_j$, the parameter $\alpha$ (the size of the
displacement) is divided by two and we go back to (ii). Otherwise,
we instead generate a new configuration from $\{\theta\}_{j+1}$.

\end{enumerate}

In the last step the reduction of $\alpha$ is limited by the
precision value required for $\epsilon$. This limit is reached
when $\left |\alpha\vec{\nabla}E_T\right |<\epsilon$.

We have used the two methods discussed above to obtain the
equilibrium positions of the magnetizations. Each method was
applied for each value of the applied magnetic field and for each
set of experimental parameters. We choose the configuration with
the lowest energy furnished by both methods as the equilibrium
configuration.

\section{Numerical results}
In this section we present the numerical results obtained for the
magnetization curves of quasiperiodic magnetic multilayers. In all
situations we have considered the cubic anisotropy effective field
$H_{ca}=0.5$ kOe which corresponds to Fe$(100)$ with $t>30$\AA. In
our calculations we have used two sets of experimental values for
the bilinear and biquadratic exchange coupling: (i) the first one
with $H_{bl}=-1.0$ kOe and $H_{bq}=0.1$ kOe. It lies in the region
of the first antiferromagnetic peak of the bilinear exchange
coupling, corresponding to a realistic sample whose Cr thickness
is about $10$\AA;  (ii) the second set with $H_{bl}=-0.035$ kOe
and $H_{bq}=0.035$ kOe. It is in the region of the second
antiferromagnetic peak of the bilinear exchange coupling,
corresponding to a realistic sample whose Cr thickness is about
$25$\AA.

\subsection{Fibonacci magnetic multilayers}
The magnetization curves for the first set of parameters of the
Fibonacci magnetic multilayers are shown in  Fig.\ 3. For the
third generation (which corresponds to the well known Fe/Cr/Fe
trilayer), in the low field region, the magnetizations are
antiparallel. As the field  increases, they continuously rotate
toward the field direction (second order phase transition) and the
saturation is reached when the external magnetic field $H\sim
1.91$ kOe. For the fifth generation there are two first order
phase transitions at $H\sim 0.71$ kOe and $H\sim 0.87$ kOe,
respectively. The saturation is reached at $H\sim 2.93$ kOe. For
the seventh generation, there are three first order phase
transitions at $H\sim 0.28$ kOe, $H\sim 0.96$ kOe and $H\sim 1.06$
kOe, respectively. The saturation is reached at $H\sim 3.03$ kOe.
For this set of parameters the majority of the transitions are of
 second order. Note that due to the different Fe layer
thickness, for the fifth and seventh generations, the
magnetization is not zero even for zero magnetic field. In Fig.\ 4
we show the results for the second set of parameters. For the
third generation, due to the strong biquadratic field, there is no
antiparallel phase in the low field region. Two magnetic phases
are present: $90^{\circ}$ ($H<72$ Oe) and saturated ($H>72$ Oe).
The fifth generation presents three magnetic phases: (i)
$90^{\circ}$ ($H<72$ Oe); (ii) almost saturated ($72$ Oe $<H<0.14$
kOe); and (iii) saturated ($H>0.14$ kOe). The seventh generation
presents four magnetic phases from $90^{\circ}$ ($H<36$ Oe) to the
saturated regime ($H>0.14$ kOe). All transitions are of first
order. Note the striking self-similar pattern shown by the
magnetization profile in this figure (see the windows).

\subsection{Double period magnetic multilayers}
Figs.\ 5a and 5b show our results for the double period  magnetic
multilayers, considering the first set of parameters. For the
second generation, due to the double Fe layer, the magnetization
has about $1/3$ of its saturation value for zero magnetic field.
There is a first order phase transition from antiparallel to an
asymmetric phase at $H\sim0.69$ kOe. In this phase, the
magnetizations are asymmetrically oriented along the magnetic
field. When $H\sim 1.34$ kOe the saturated phase emerges. For the
fourth generation, for zero magnetic field, the magnetization has
about $10\%$ of its saturation value due to the different
thickness of the Fe layers. There is a first order phase
transition at $H\sim 0.29$ kOe and the saturation is reached at
$H\sim 3.27$ kOe. All other phase transitions are of second order.
For the second set of parameters (see Fig.\ 6), on the contrary,
all transitions are of first order. For this set of parameters,
there is no antiparalell phase in the low field region, due to the
strong biquadratic field. For the second generation, the
magnetization is about $2/3$ of its saturation value when $H=0$.
There are two magnetic phases: (i) $90^{\circ}$ ($0 < H < 72$ Oe)
and (ii) saturated ($H> 72$ Oe). For the fourth generation, the
magnetization for zero magnetic field  is about $1/2$ of its
saturation value. Four magnetic phases are present, from the
$90^{\circ}$ ($0<H<38$ Oe) to the saturated phase ($H>0.14$ kOe).
As in the Fibonacci case,  a self-similar pattern is also present
in the magnetization curves (see the window).

\section{Conclusions}
We have studied quasiperiodic magnetic multilayers, composed by
ferromagnetic Fe layers separated by nonmagnetic Cr layers,
arranged according to the Fibonacci and double period
quasiperiodic sequences. We consider that the Fe layers are linked
by bilinear and biquadratic exchange couplings through Cr layers
and present cubic anisotropy. The external magnetic field is
applied in the plane of the layers and along an easy axis. We have
used two numerical methods to determine the equilibrium
configurations of the layers's  magnetizations.  The magnetization
curves of these artificial structures were calculated considering
two sets of experimental parameters recently reported
\cite{6,7,8}. Our results show that quasiperiodic magnetic
multilayers exhibit a rich variety of configurations induced by
the external magnetic field. In particular two points may be
emphasized: (i) the effect of different thickness of Fe layers and
(ii) the effect of the biquadratic exchange coupling.

The effect of different thickness of Fe layers is evident in the
low field region. In that region, due to these differences, there
is a net magnetization even if the alignment is antiparallel and
the external magnetic field is zero. Besides, the nature of the
phase transitions are changed by the different thickness (Fig.\ 3a
shows only second order phase transitions, while Fig.\ 5a presents
an additional first order phase transition). These results suggest
that, varying the thickness of Fe layers, it is possible to tailor
magnetic multilayers to present desired specific phase transitions
and critical fields. However, as the thickness of Fe layers
increases, the crystalline anisotropy of Fe (100) films on Cr
(100) also increases. Fortunately, as a characteristic of the
quasiperiodic multilayers arrangements considered here, the
maximum number of joint Fe layers is two (for the Fibonacci case)
and three (for the double period case), no matter is the value of
their generation numbers. Besides, from a thickness greater than
$40$\AA, the crystalline anisotropy reaches saturation \cite{7}.

On the other hand, the biquadratic exchange coupling plays a
remarkable role in the features of the magnetization curves. For
example, when the bilinear exchange coupling prevails, the
majority of the transitions are of second order character (see
Figs.\ 3 and 5). However, when the biquadratic exchange is
compared to the bilinear one, in the presence of a stronger
crystalline anisotropy \cite{8}, the transitions are characterized
by discontinuous jumps in the magnetization that indicate first
order phase transitions. This can be considered as the basic
signature of the biquadratic exchange coupling (see Figs.\ 4 and
6), although for the case where there is no biquadratic term, a
first order phase transition appears due to the anisotropy
\cite{24}. Furthermore, as shown by the windows in these figures,
the magnetization curves of higher generations reproduce the
magnetization curves of lower generations. This self-similar
behavior is a general characteristic of quasiperiodic systems,
although it is not present when the bilinear exchange prevails
(Figs.\ 3 and 5). A possible explanation for these different
behaviors is because the biquadratic exchange coupling induces
long range correlations that emphasize the quasiperiodicity of the
system. These long range correlations make the whole structure
{\it seeing} its quasiperiodicity, which is reflected in the
magnetization curves. This argument is reinforced by previous
works on the correlation lengths of magnetic systems presenting
biquadratic exchange coupling (see, for example, S\o rensen and
Young\cite{24}).

The most appropriate experimental technique for studying the
magnetization curves of magnetic films is the magneto-optical Kerr
effect (MOKE)\cite{8}. However, because the MOKE measurements
provide surface sensitivity on the scale of the optical
penetration depth ($\sim 10$ \AA), it is necessary to use also a
superconductor quantum interface device (SQUID) magnetometry
\cite{19}. The two techniques prove complementary in understanding
the switching behavior of the multilayer films, as far as the
magnetization curves are concern. We hope that the present results
can stimulate experimental studies of these structures.

\vskip 0.5 cm

\noindent {\it Acknowledgements}: We would like to thank the
Brazilian Research Council CNPq for financial support and CESUP-RS
where part of the numerical calculation was done.

\newpage
\centerline
{\bf Figure Captions}
\begin{enumerate}

\item The third and fifth Fibonacci generations and their magnetic
counterpart.

\item Same as Fig. 1 for the second and fourth double period generations.

\item Magnetization versus applied field for the third (a), fifth (b) and
seventh (c) Fibonacci generations with $|H_{bq}|/|H_{bl}|=0.10$,
corresponding to a realistic sample whose Cr thickness is about
$10$\AA. We have considered the cubic anisotropy effective field
$H_{ca}=0.5$ kOe, which corresponds to Fe$(100)$ with $t>30$\AA.

\item Same as Fig. 3, but for $|H_{bq}|/|H_{bl}|=1.0$, corresponding to a
realistic sample whose Cr thickness is about $25$\AA.

\item Magnetization versus applied field for the second (a) and fourth (b)
double period generations with $|H_{bq}|/|H_{bl}|=0.10$. The cubic
anisotropy effective field is again $H_{ca}=0.5$ kOe. The Cr
thickness is $10$\AA.

\item Same as Fig. 5, but for $|H_{bq}|/|H_{bl}|=1.0$, corresponding to a
realistic sample whose Cr thickness is about $25$\AA.

\end{enumerate}

\end{document}